\documentclass[twocolumn,showpacs,preprintnumbers,amsmath,amssymb]{revtex4}

\usepackage{graphicx}
\usepackage{dcolumn}
\usepackage{bm}
\usepackage{color}

\begin{document}


\title{Scaling Laws in Human Language}

\author{Linyuan L\"{u}}
\author{Zi-Ke Zhang}
\author{Tao Zhou}
\email{zhutou@ustc.edu} \affiliation{Web Sciences Center, University of Electronic Science and Technology of China, Chengdu 610054, People's Republic of China \\Department of Physics, University of Fribourg, Chemin du Muse, Fribourg 1700, Switzerland \\ Beijing Computational Science Research Center, Beijing 100084,  People's Republic of China}

\date{\today}

\begin{abstract}
Zipf's law on word frequency is observed in English, French, Spanish, Italian, and so on, yet it does not hold for Chinese, Japanese or Korean characters. A model for writing process is proposed to explain the above difference, which takes into account the effects of finite vocabulary size. Experiments, simulations and analytical solution agree well with each other. The results show that the frequency distribution follows a power law with exponent being equal to 1, at which the corresponding Zipf's exponent diverges. Actually, the distribution obeys exponential form in the Zipf's plot. Deviating from the Heaps' law, the number of distinct words grows with the text length in three stages: It grows linearly in the beginning, then turns to a logarithmical form, and eventually saturates. This work refines previous understanding about Zipf's law and Heaps' law in language systems.
\end{abstract}

\pacs{89.20.Hh, 89.75.Hc}

\maketitle

Uncovering the statistics and dynamics of human language helps in characterizing the universality, specificity and evolution of cultures \cite{Hawkins1992,Caplan1994,Lightfoot1999,Nowak1999,Cancho2001,Nowak2002,Hauser2002,Abrams2003,Lieberman2007,Lambiotte2007,Petersen2011}. Two scaling relations, Zipf's law \cite{Zipf1949} and Heaps' law \cite{Heaps1978}, have attracted much attention from academic community. Denote $r$ the rank of a word according to its frequency $Z(r)$, Zipf's law is the relation $Z(r)\sim{r}^{-\alpha}$, with $\alpha$ being the Zipf's exponent. Zipf's law was observed in many human languages, including English, French, Spanish, Italian, and so on \cite{Zipf1949,Kanter1995,Cancho2002}. Heaps' law is formulated as $N_t\sim t^\lambda$, where $N_t$ is the number of distinct words when the text length is $t$, and $\lambda\leq 1$ is the so-called Heaps' exponent. These two laws coexists in many language systems. Gelbukh and Sidorov \cite{Gelbukh2001} observed these two laws in English, Russian and Spanish texts, with different exponents depending on languages. Similar results were recently reported for the corpus of web texts \cite{Serrano2009}, including the\emph{ Industry Sector database}, the \emph{Open Directory} and the \emph{English Wikipedia}. The occurrences of tags for online resources \cite{Cattuto2007,Cattuto2009}, keywords for scientific publications \cite{Zhang2008} and words contained by web pages resulted from web searching \cite{Lansey2009} also simultaneously display the Zipf's law and Heaps' law. Interestingly, even the identifiers in programs by Java, C++ and C languages exhibit the same scaling laws \cite{Zhang2009}.

The Zipf's law in language systems could result from a rich-get-richer mechanism as suggested by the Yule-Simon model \cite{Simon1955,Simkin2011}, where a new word is added to a text with probability $q$ and an appeared word is randomly chosen and copied with probability $1-q$. A word appears more frequently thus has high probability to be copied, leading to a power-law word frequency distribution $p(k)\sim{k}^{-\beta}$ with $\beta=1+1/(1-q)$. Dorogovtsev and Mendes modeled the language processing as evolution of a word web with preferential attachment \cite{Dorogovtsev2001}. Zanette and Montemurro \cite{Zanette2005} as well as Cattuto \emph{et al.} \cite{Cattuto2006} accounted for the memory effects, say the recently used words have higher probability to be chosen than the words occurred long time ago. These works can be considered as variants of the Yule-Simon model. Meanwhile, the Heaps' law may originate from the memory and bursty nature of human language \cite{Ebeling1994,Kleinberg2003,Altmann2009}.

\begin{table}
\caption{The basic statistics of the four books. $\beta$ is the exponent of the
power-law frequency distribution and $N_T$ is the total number of distinct characters. }
\begin{center}
\begin{tabular} {cccccc}
  \hline \hline
   Books     & $V$  &  $N_T$ & $k_{max}$ & $k_{min}$ & $\beta$  \\
   \hline
   The Story of the Stone & 727601 & 4239  & 21054 & 1 & 1.09 \\
   The Battle Wizard & 1020336 & 4178 & 20028 & 1 & 1.03 \\
   Into the White Night & 420935 & 2182 & 18992 & 1 & 1.00 \\
   History of the Three Kingdoms & 157201 & 1139 & 5929 & 1 & 1.07 \\
   \hline \hline
    \end{tabular}\label{statistical}
\end{center}
\end{table}

\begin{figure*}
\begin{center}
\includegraphics[width=4.00cm]{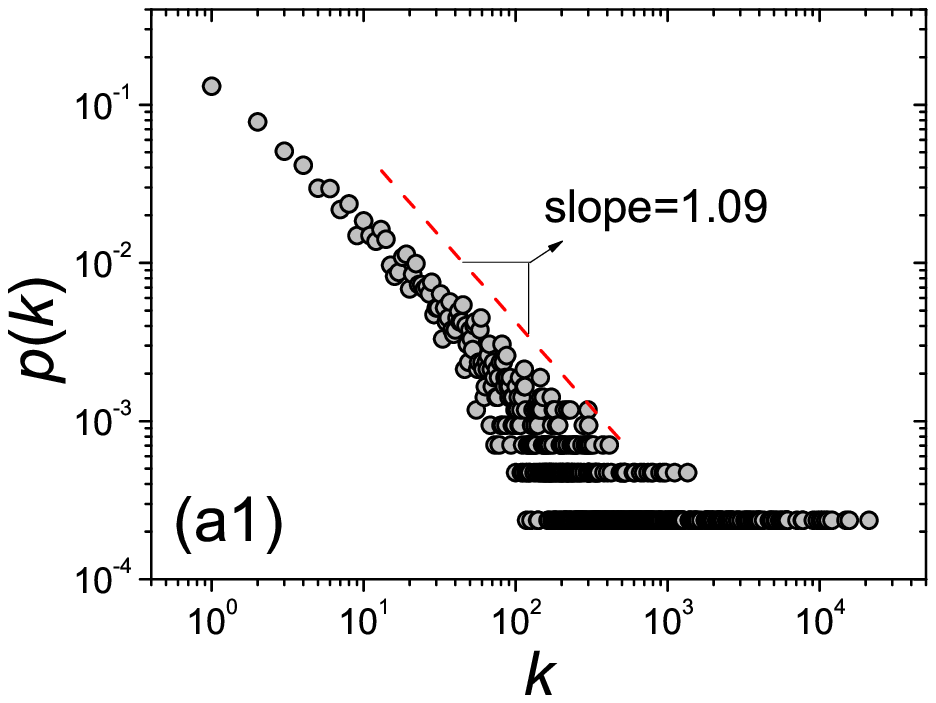}
\includegraphics[width=4.10cm]{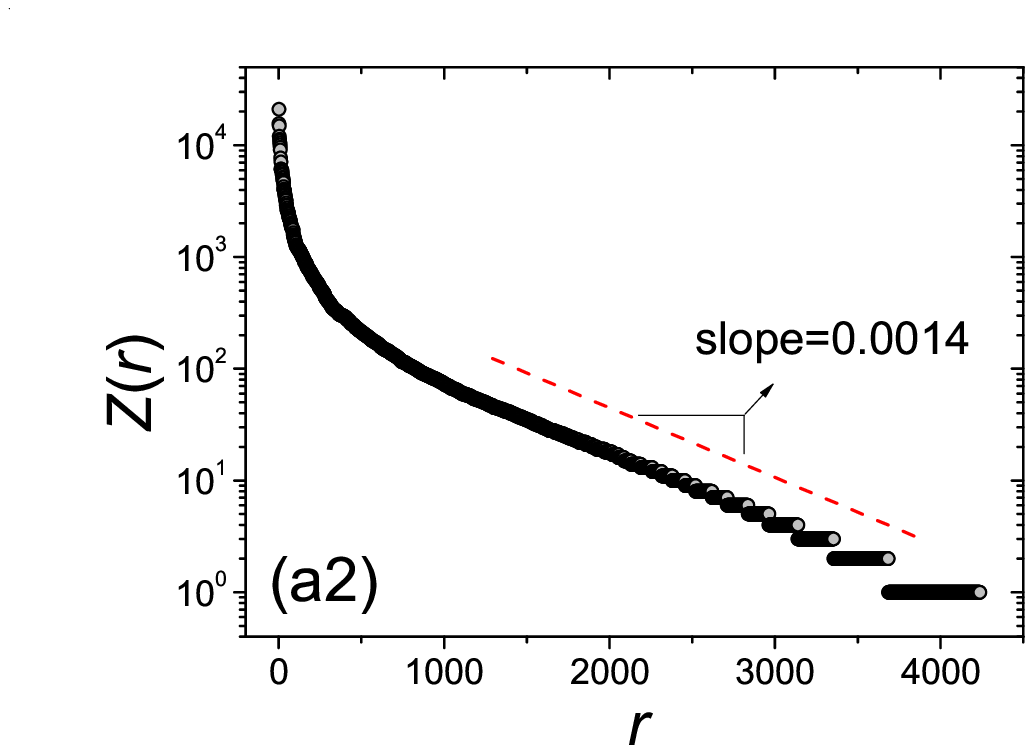}
\includegraphics[width=4.25cm]{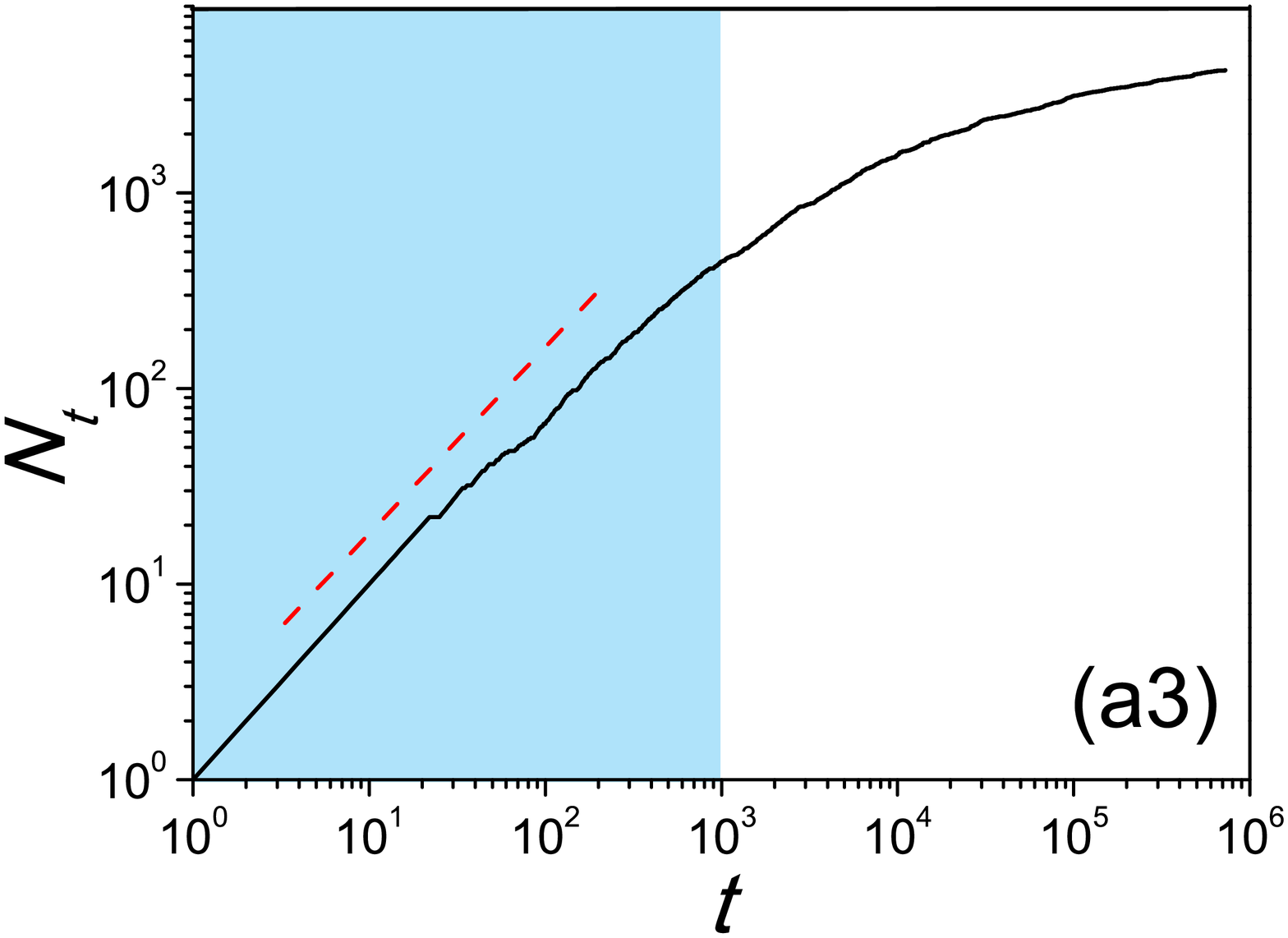}
\includegraphics[width=4.25cm]{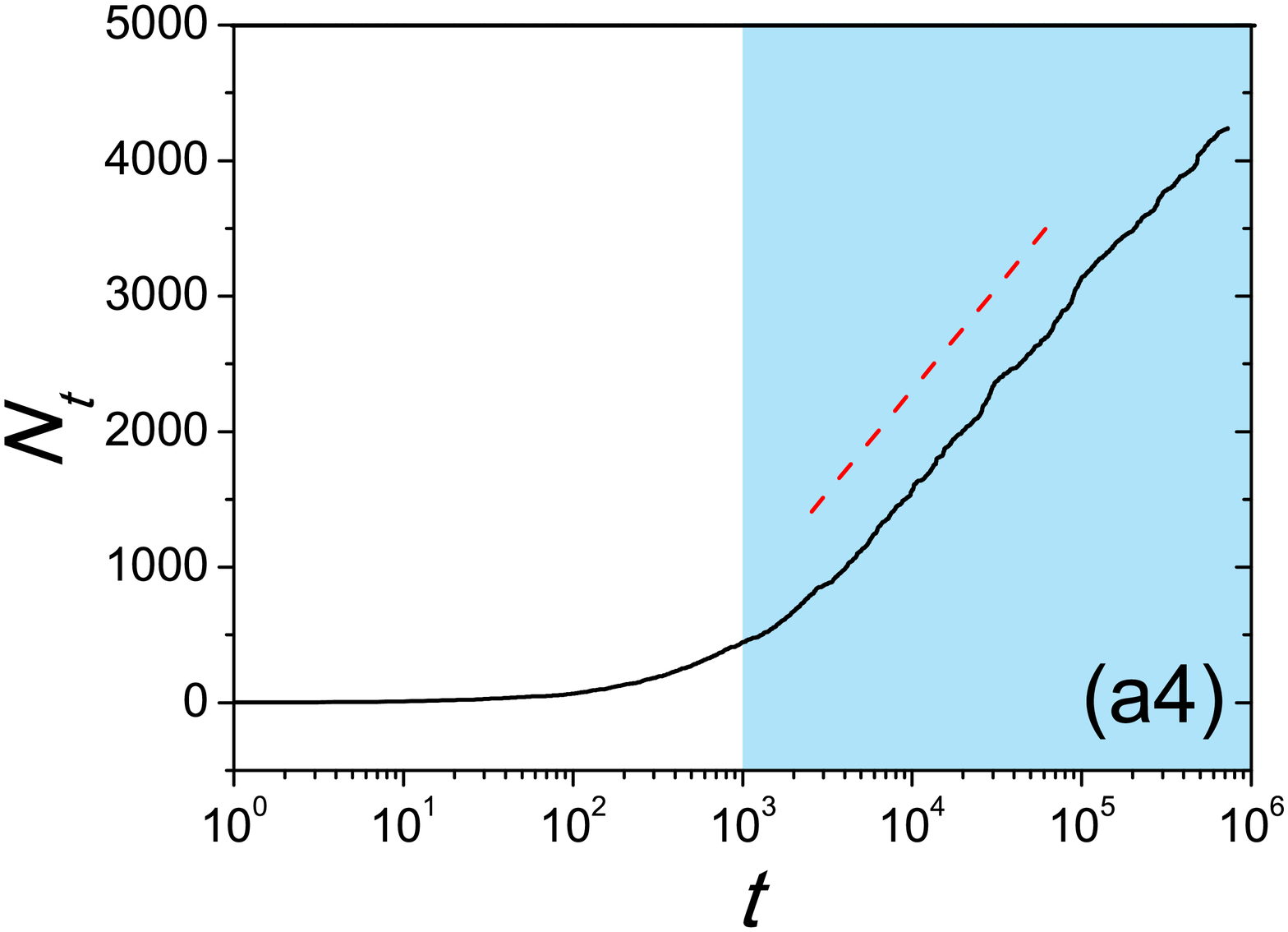}
\includegraphics[width=4.10cm]{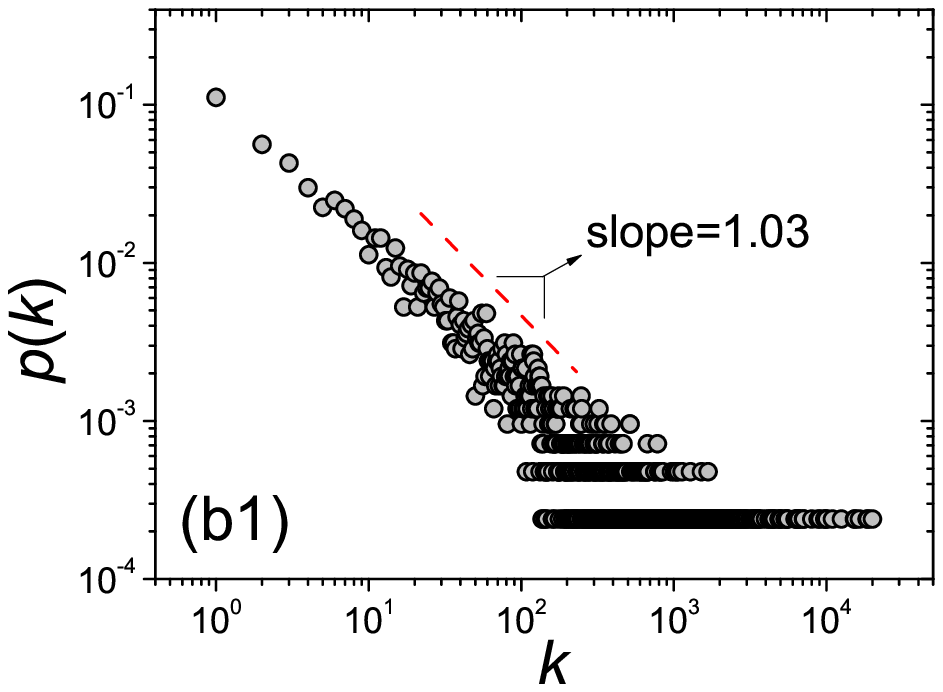}
\includegraphics[width=4.10cm]{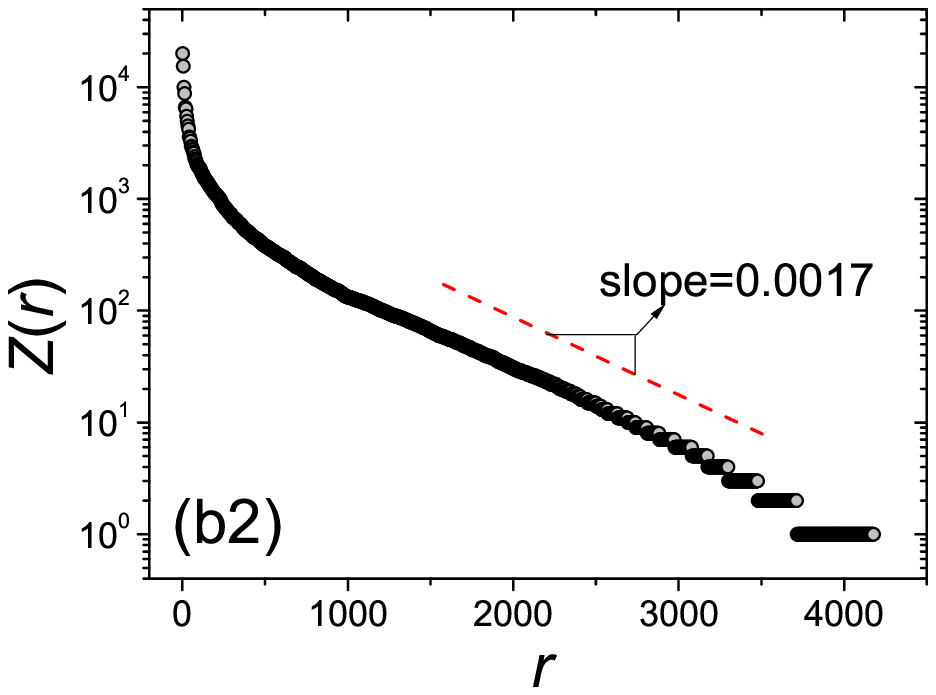}
\includegraphics[width=4.25cm]{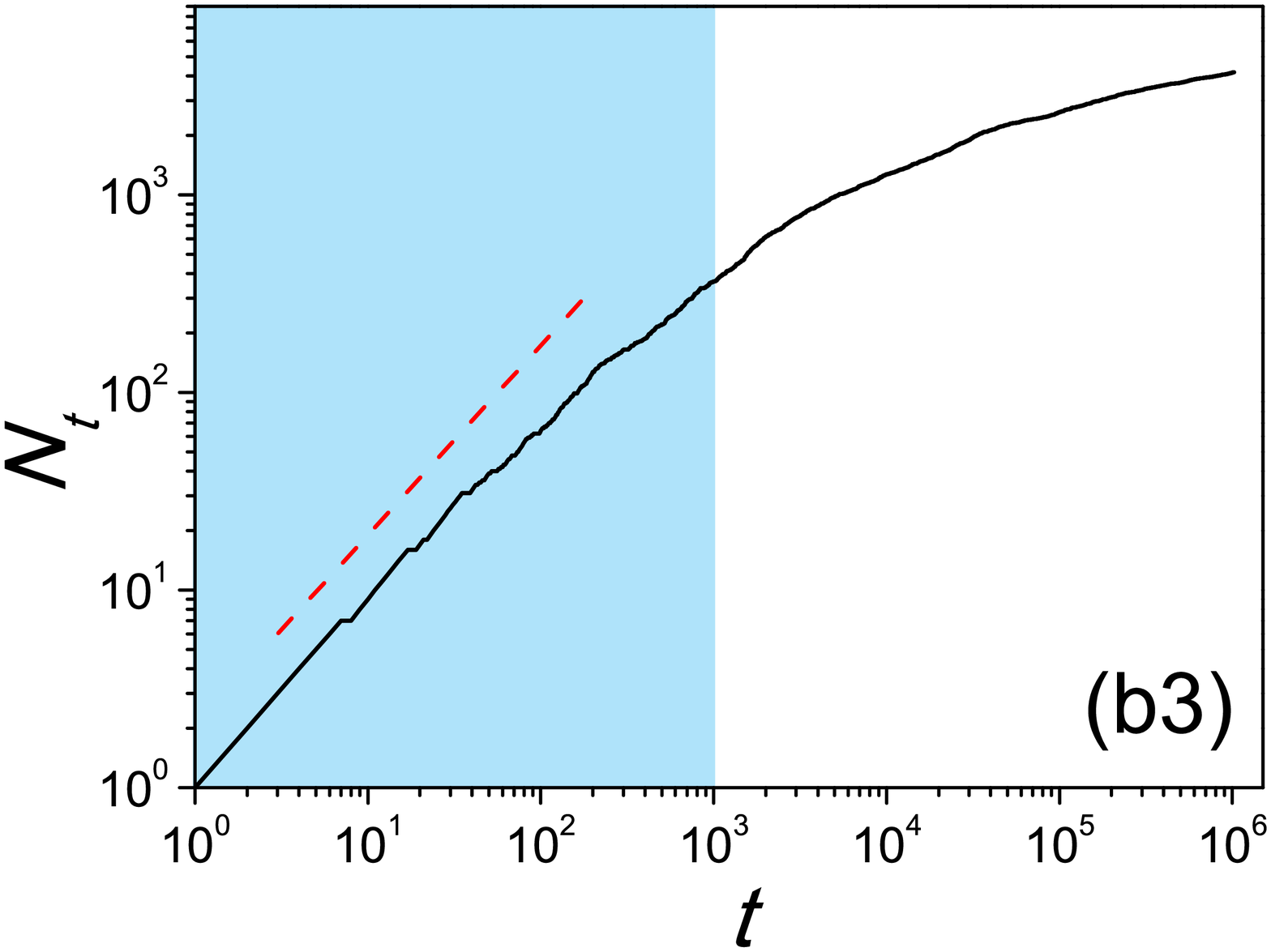}
\includegraphics[width=4.25cm]{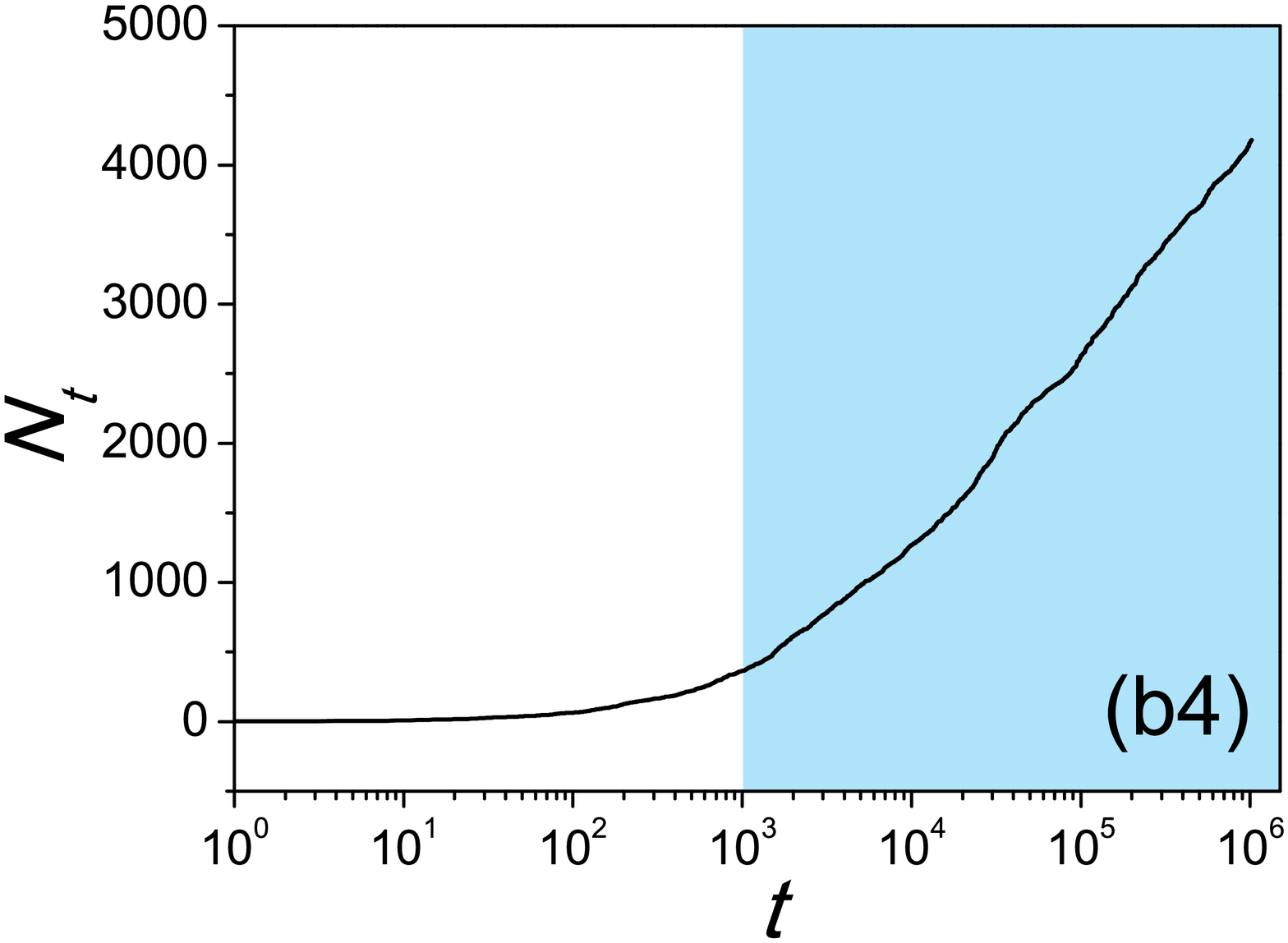}
\includegraphics[width=4.10cm]{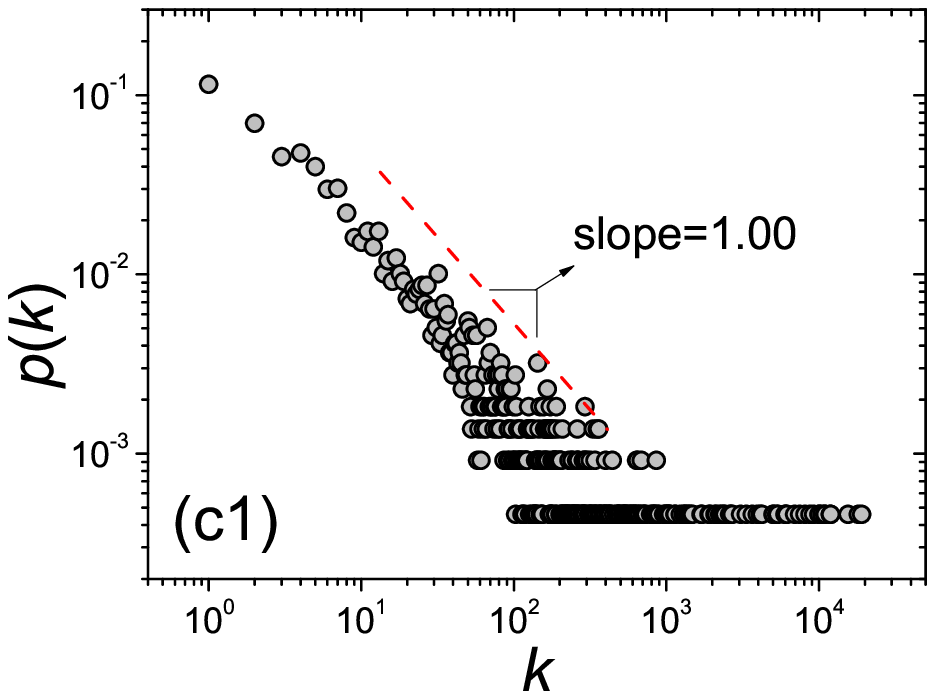}
\includegraphics[width=4.10cm]{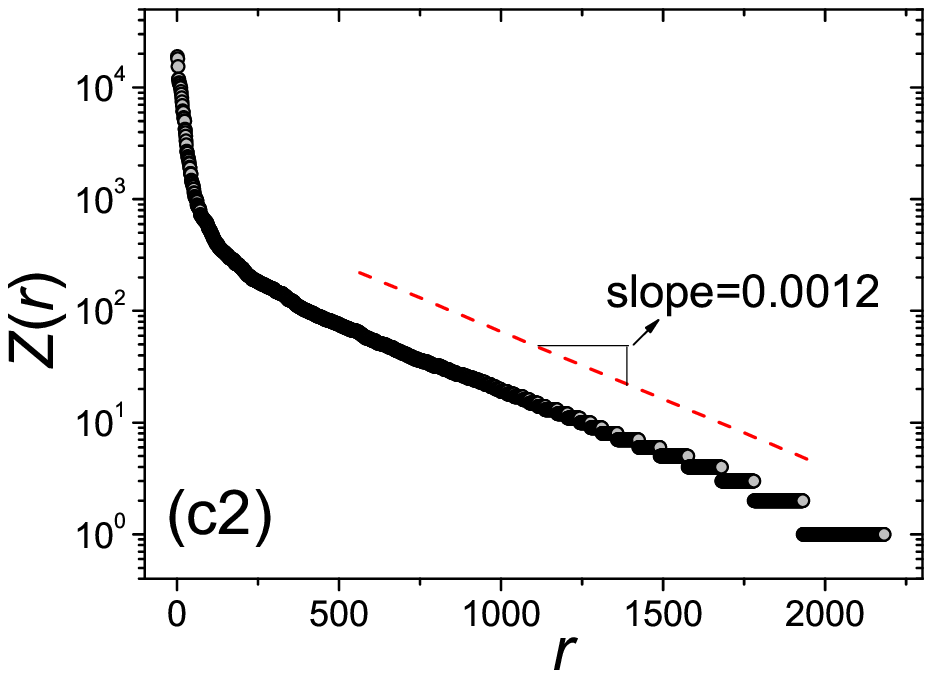}
\includegraphics[width=4.25cm]{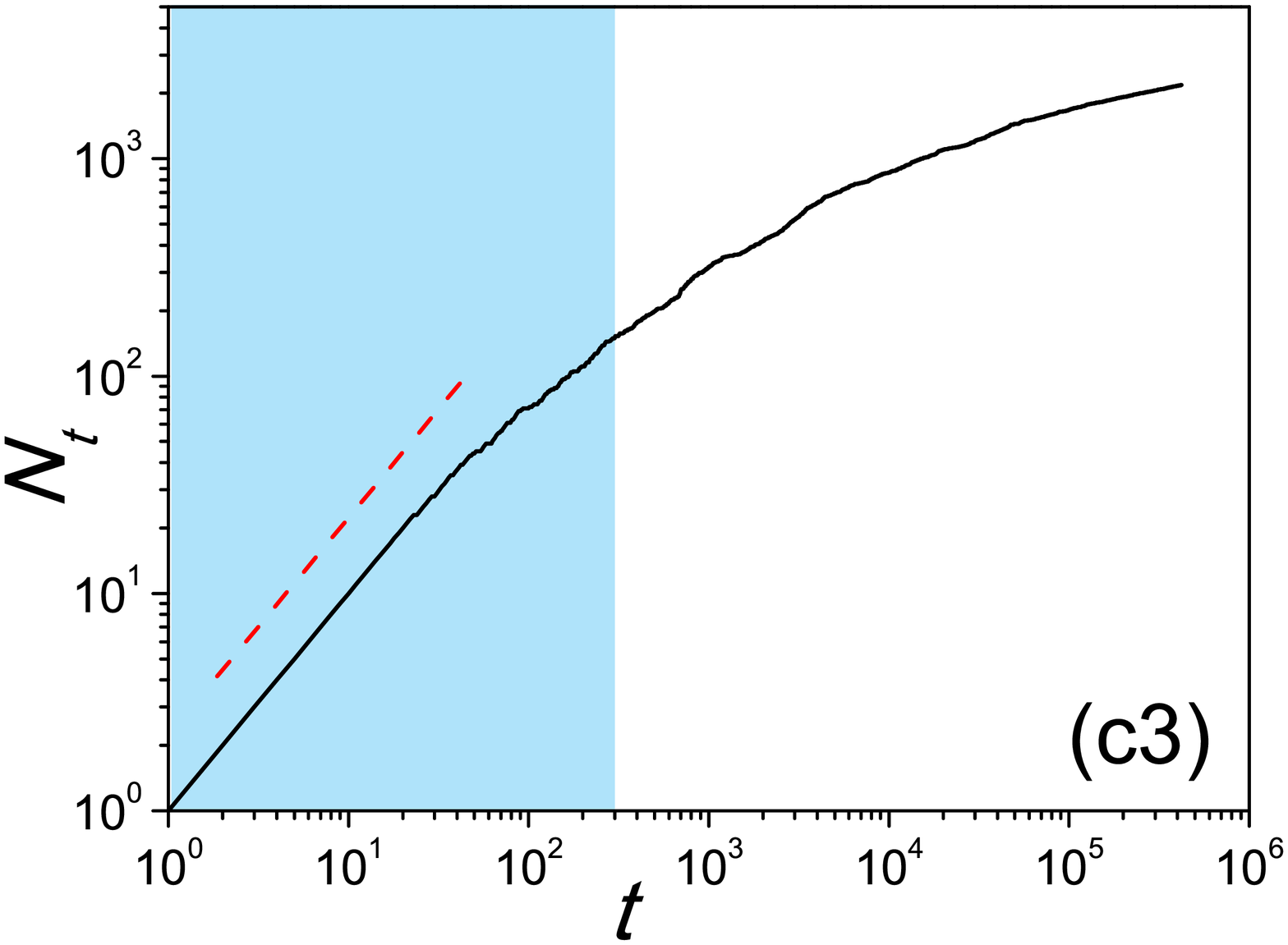}
\includegraphics[width=4.25cm]{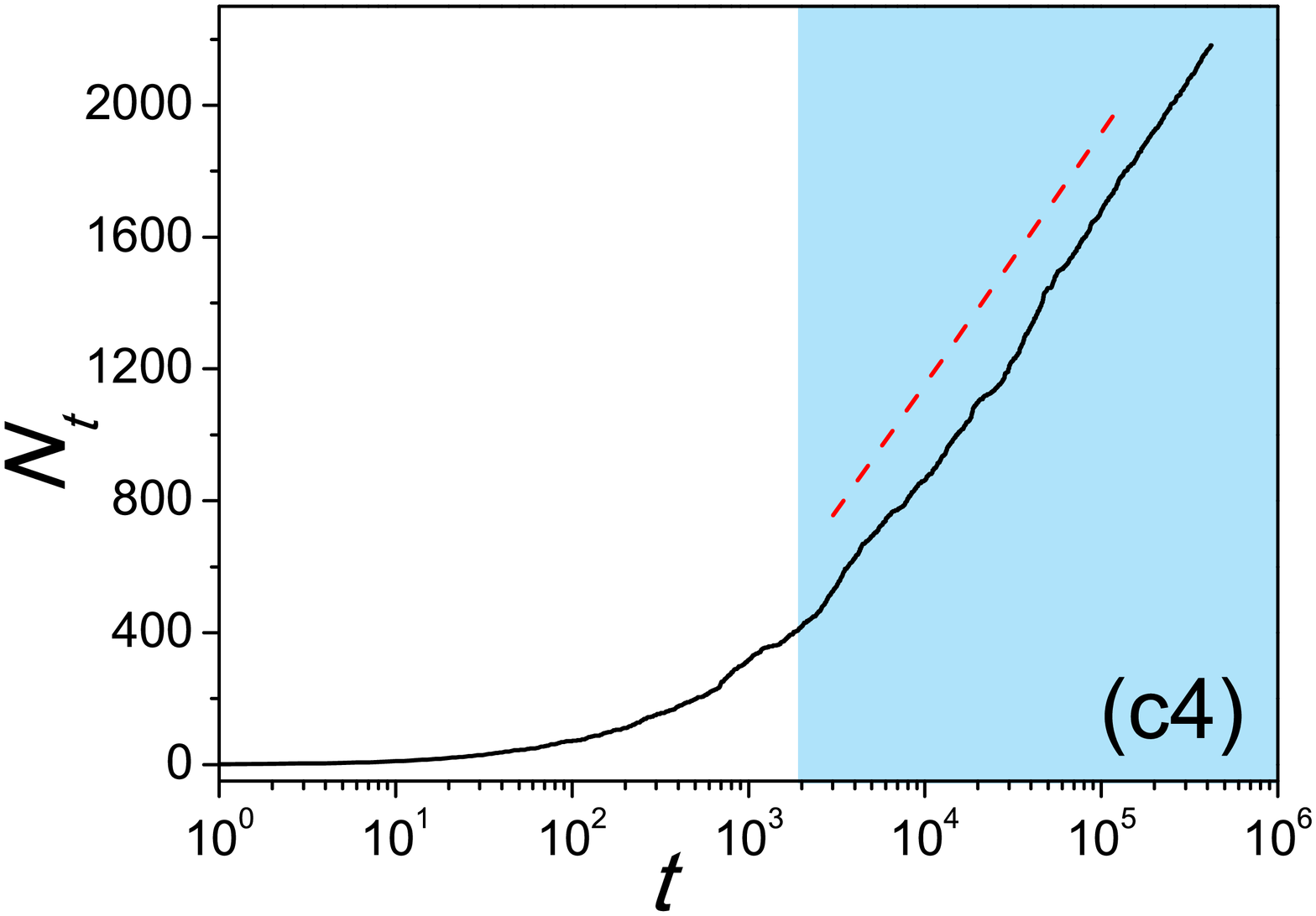}
\includegraphics[width=4.10cm]{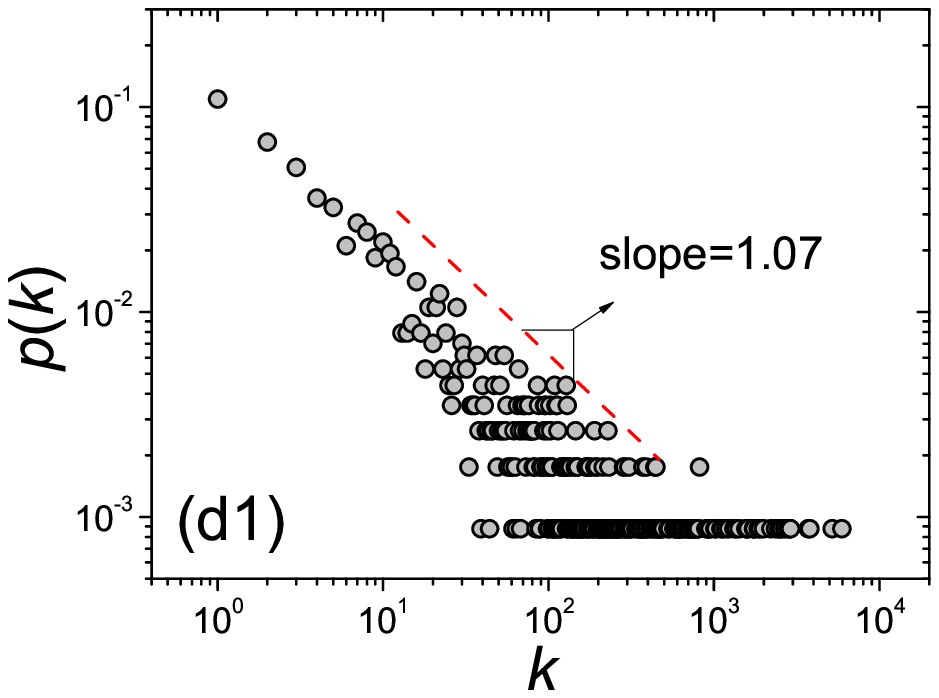}
\includegraphics[width=4.10cm]{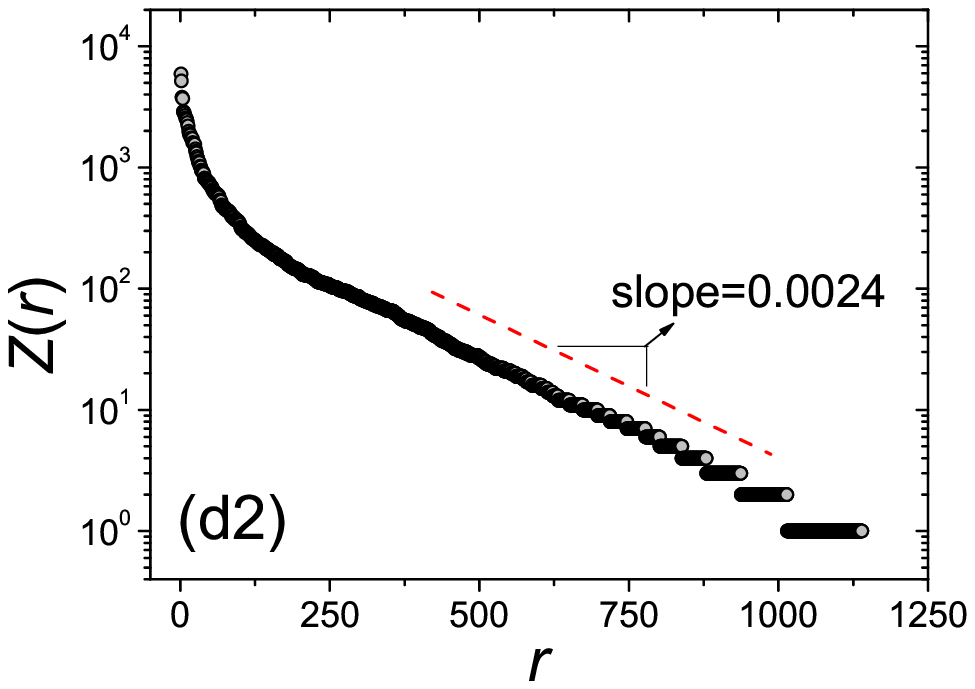}
\includegraphics[width=4.25cm]{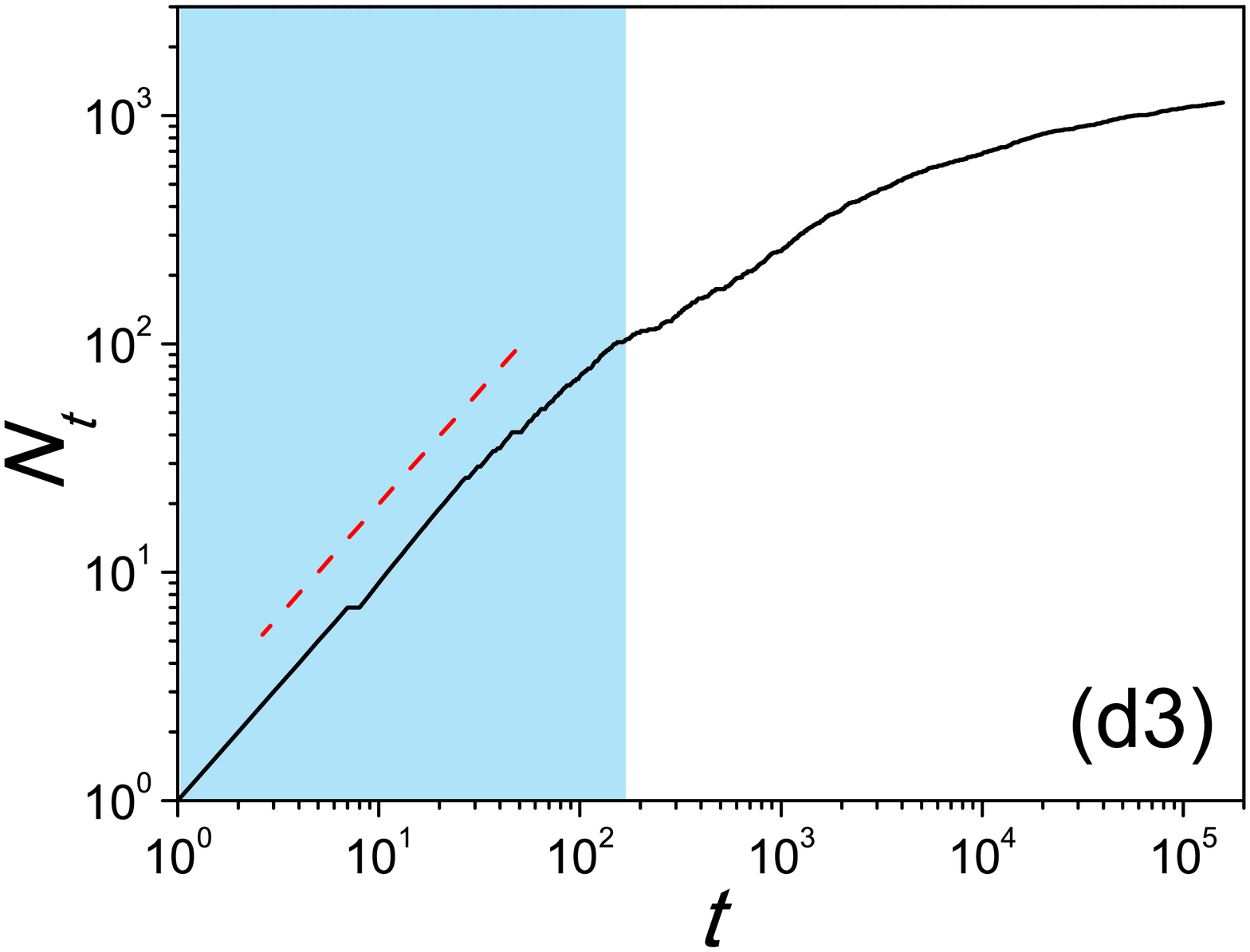}
\includegraphics[width=4.25cm]{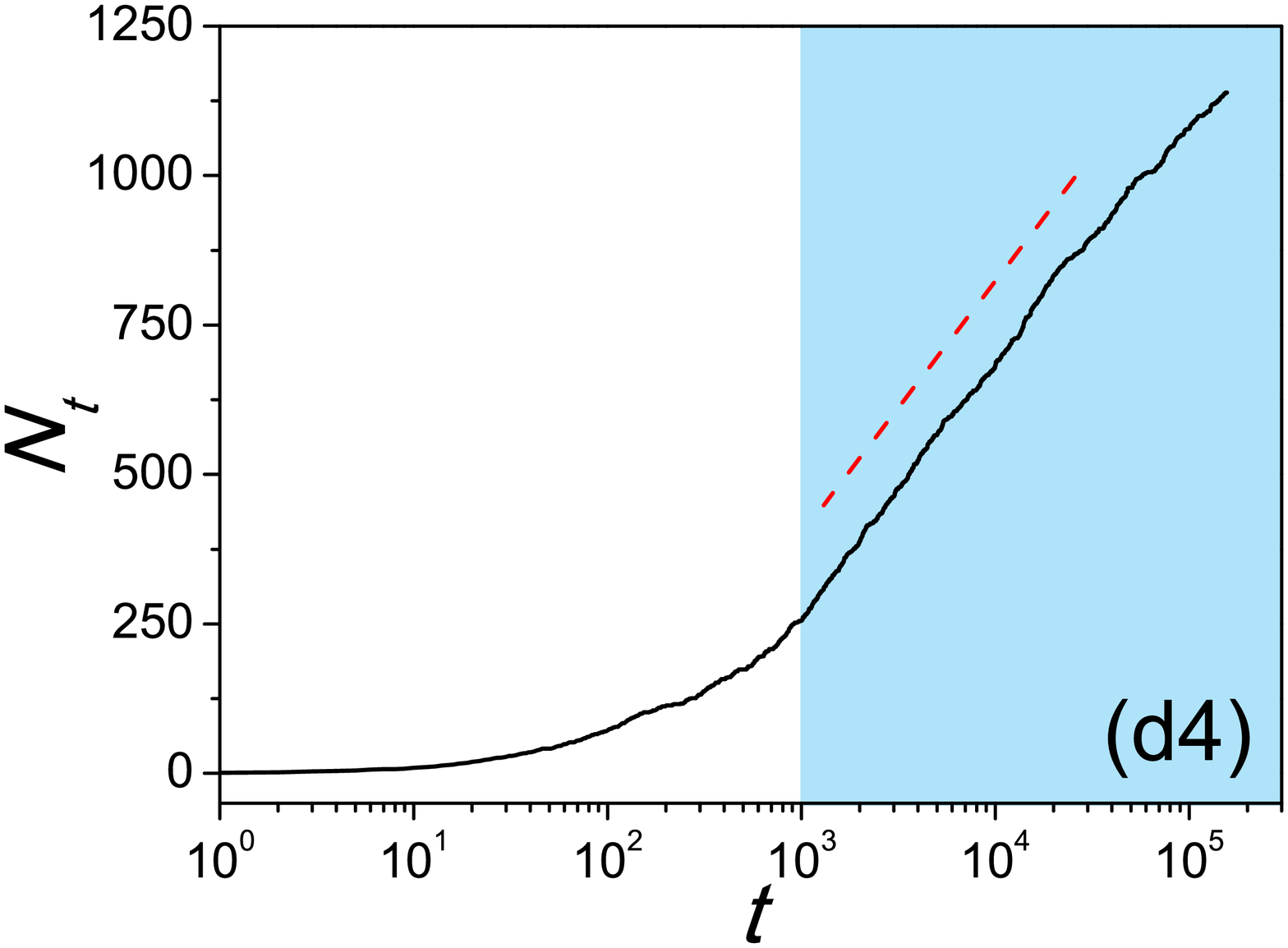}
\caption{(Color online) The character frequency distribution of \emph{The Story of the Stone}: (a1) $p(k)$ with log-log scale and (a2) $Z(r)$ with log-linear scale. The number of distinct words versus the text length of \emph{The Story of the Stone} in (a3) log-log scale and (a4) linear-log scale. Similar plots in (b1-b4), (c1-c4) and (d1-d4) are for the books \emph{The Battle Wizard}, \emph{Into the White Night} and \emph{The History of the Three Kingdoms}, respectively. The power-law exponent $\beta$ is obtained by using the maximum likelihood estimation \cite{Goldstein2004,Clauset2009}, while the exponent in the Zipf's plot is obtained by the least square method excluding the head (i.e., $r>500$ for Chinese books and $r>200$ for Japanese and Korean books). }\label{empiricalHLM}
\end{center}
\end{figure*}

Real language systems to some extent deviate from these two scaling laws and display more complicated statistical regularities. Wang \emph{et al.} \cite{Wang2005} analyzed representative publications in Chinese, and showed that the character frequency distribution exhibits an exponential feature. L\"u \emph{et al.} \cite{Lu2010} pointed out that in a growing system, if the appearing frequencies of elements obey the Zipf's law with stable exponent, then the number of distinct elements grows in a complicated way with the Heaps' law only an asymptotical approximation. This deviation from the Heaps' law was further emphasized and mathematically proved by Eliazar \cite{Eliazar2011}. Empirical analyses on real language systems showed similar deviation \cite{Bernhardsson2009}. Via extensive analysis on individual Chinese, Japanese and Korean books, as well as a collection of more than $5\times 10^4$ Chinese books, we found even more complicated phenomena: (i) the character frequency distribution follows a power law yet it decays exponentially in the Zipf's plot; (ii) with the increasing of text length, the number of distinct characters grows in three different stages: linear, logarithmical and saturated. All these unreported regularities may result from the finite vocabulary size, which is further verified by a simple theoretical model.


We first show some experimental results about the statistical regularities on Chinese, Japanese and Korean literatures, which are typical examples generated from a vocabulary of very limited size if we look at the character level. There are in total more than $9\times 10^4$ Chinese characters, yet only 3000 to 4000 of which are used frequently (Taiwan and Hong Kong respectively identify 4808 and 4759 frequently used characters, while mainland China has two versions of the list of frequently used characters, one contains 2500 characters and the other contains 3500 characters), and the number of Japanese and Korean characters are even smaller. We start with four famous books, the first two are in Chinese, the third one is in Japanese and the last one is in Korean: (i) \emph{The Story of the Stone} (aliases: \emph{The Dream of the Red Chamber}, \emph{A dream of Red Mansions} and \emph{Hong Lou Meng}), written by Xueqin Cao in the mid-eighteenth century during the reign of Emperor Chien-lung of the Qing Dynasty; (ii) \emph{The Battle Wizard} (aliases: \emph{Tian Long Ba Bu} and
\emph{Demi-Gods and Semi-Devils}), a kung fu novel written by Yong Jin; (iii) \emph{Into the White Night}, a modern novel written by Higashino Keigo; (iv) \emph{The History of the Three Kingdoms}, a very famous history book by Shou Chen in China and then translated into Korean. These books cover disparate topics and types and were accomplished in far different dates. The basic statistics of these books are presented in Table 1.

Figure 1 reports the character frequency distribution $p(k)$, the Zipf's plot on character frequency $Z(r)$ and the growth of the number of distinct characters $N_t$ versus the total number of characters appeared in the text. As shown in figure 1, the character frequency distributions are power-law, meanwhile the frequency decays exponentially in the Zipf's plot, which is in conflict to the common sense that a power-law probability density function always corresponds to a power-law decay in the Zipf's plot. Actually, there exists a relation between two exponents $\alpha$ and $\beta$ as $\alpha=\frac{1}{\beta-1}$ \cite{Lu2010}, and thus when $\beta$ gets close to 1, the exponent $\alpha$ will diverge and thus the decaying function in Zipf's plot could not be well characterized by a power law. Therefore, if we observe a non-power-law decaying in the Zipf's plot, we cannot immediately deduce that the corresponding probability density function is not a power law -- it is possibly a power law with exponent close to 1. Note that, in the Zipf's plots, the turned-up head contains a few hundreds of characters, majority of which play the similar role to the auxiliary words, conjunctions or prepositions in English.

Figure 1 also indicates that the growth of distinct characters cannot be described by the Heaps' law. Indeed, there are two distinguishable stages: In the early stage, $N_t$ grows approximately linearly with the text length $t$, and in the later stage, $N_t$ grows logarithmically with $t$. Figure 3 presents the growth of distinct characters for a large collection of 57755 Chinese books consisting of about $3.4\times 10^9$ characters and 12800 distinct characters. In addition to those observed in figure 1 and figure 2, $N_t$ displays a strongly saturated behavior when the text length $t$ is much bigger than the total distinct characters in the vocabulary. In summary, the experiments on Chinese, Japanese and Korean literature show us some unreported phenomena: the character frequency obeys a power law with exponent close to 1 yet it decays exponentially in the Zipf's plot, and the number of distinct characters grows in three distinguishable stages. We next propose a theoretical model to explain these observations.

\begin{figure}
\begin{center}
\includegraphics[width=4.25cm]{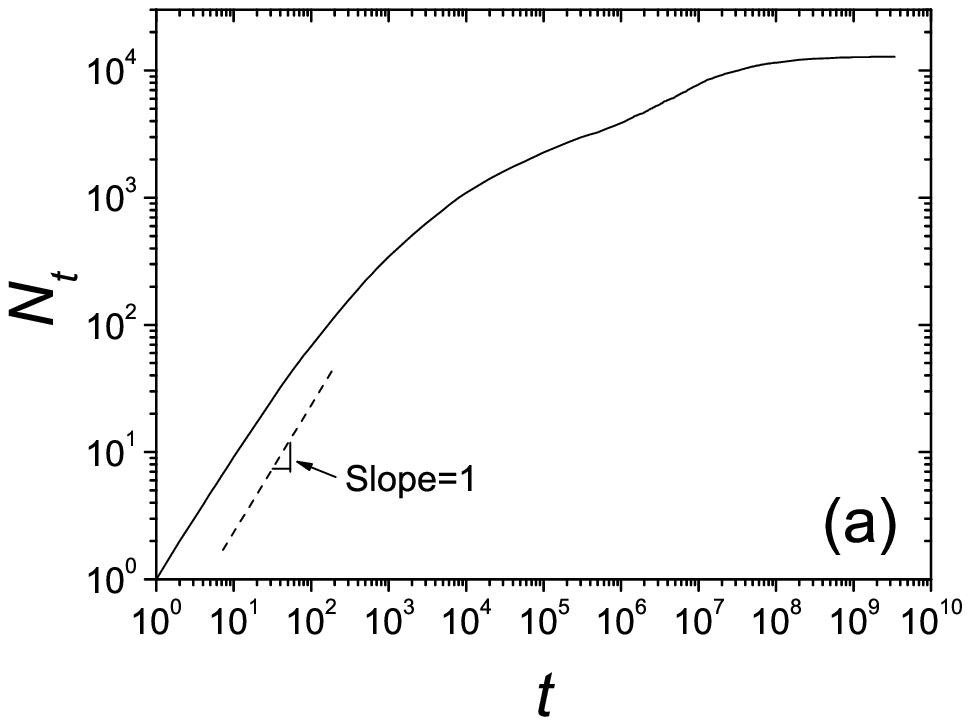}
\includegraphics[width=4.25cm]{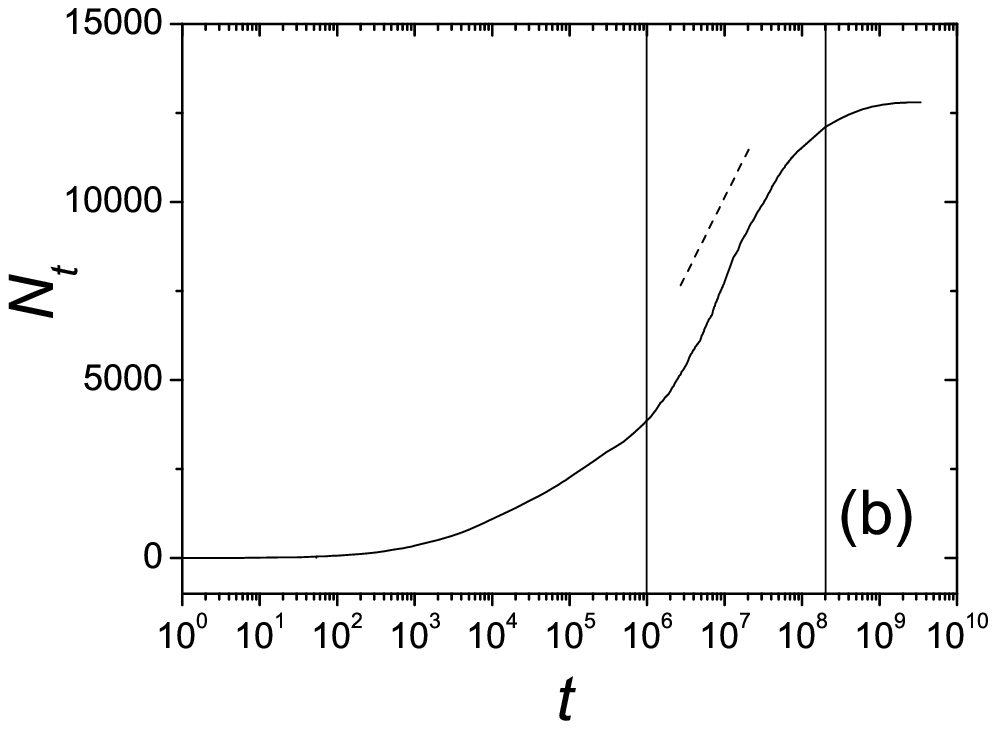}
\caption{(Color online) The growth of distinct characters in the collection of 57755 Chinese
books. The result is obtained by averaging over 100 randomly determined
orders of these books.}\label{sgrowth}
\end{center}
\end{figure}

Consider a vocabulary with finite number, $V$, of distinct characters or words. At each time step, one character will be selected from the vocabulary to form the text. Motivated by the rich-get-richer mechanism of the Yule-Simon model, at time step $t$, if the character $i$ has been used $k_{i}$ times, it will be selected with the probability
\begin{equation}
\label{fi}f(k_i)=\frac{k_i+\varepsilon}{\sum_{j=1}^{V}(k_{j}+\varepsilon)}=\frac{k_{i}+\varepsilon}{V\varepsilon+t-1},
\end{equation}
where $\varepsilon$ is the initial attractiveness of each character. Assuming that at time $t$, there are $N_t$ distinct characters in the text, and we first investigate the dependence of $N_t$ on the text length $t$. The selection at time $t+1$ can be equivalently divided into two complementary yet repulsive actions: (i) to select a character from the original vocabulary with probability proportional to $\varepsilon$, or (ii) to select a character from the $N_t$ words in the created text with probability proportional to its appeared frequency. Therefore the probability to choose a character from the original vocabulary is $\frac{V\varepsilon}{V\varepsilon+t}$, whereas $\frac{t}{V\varepsilon+t}$ from the created text. A character chosen from the created text is always old, while a character chosen from the vocabulary could be new with probability $1-\frac{N_{t}}{V}$. Accordingly, the probability that a new character appears at the $t+1$ time step, namely the growing rate of $N_t$, is
\begin{equation}
\label{Nt}\frac{dN_{t}}{dt}=\frac{V\varepsilon}{V\varepsilon+t}\left(1-\frac{N_{t}}{V}\right).
\end{equation}
With the boundary conditions $N_0=0$ and $N_{\infty}=V$, we derive
the solution of Eq.~\ref{Nt} as
\begin{equation}
\label{Nt2}N_t=V\left[1-\left(\frac{V\varepsilon}{V\varepsilon+t}\right)^{\varepsilon}\right].
\end{equation}

\begin{figure}
\begin{center}
\includegraphics[width=4.25cm]{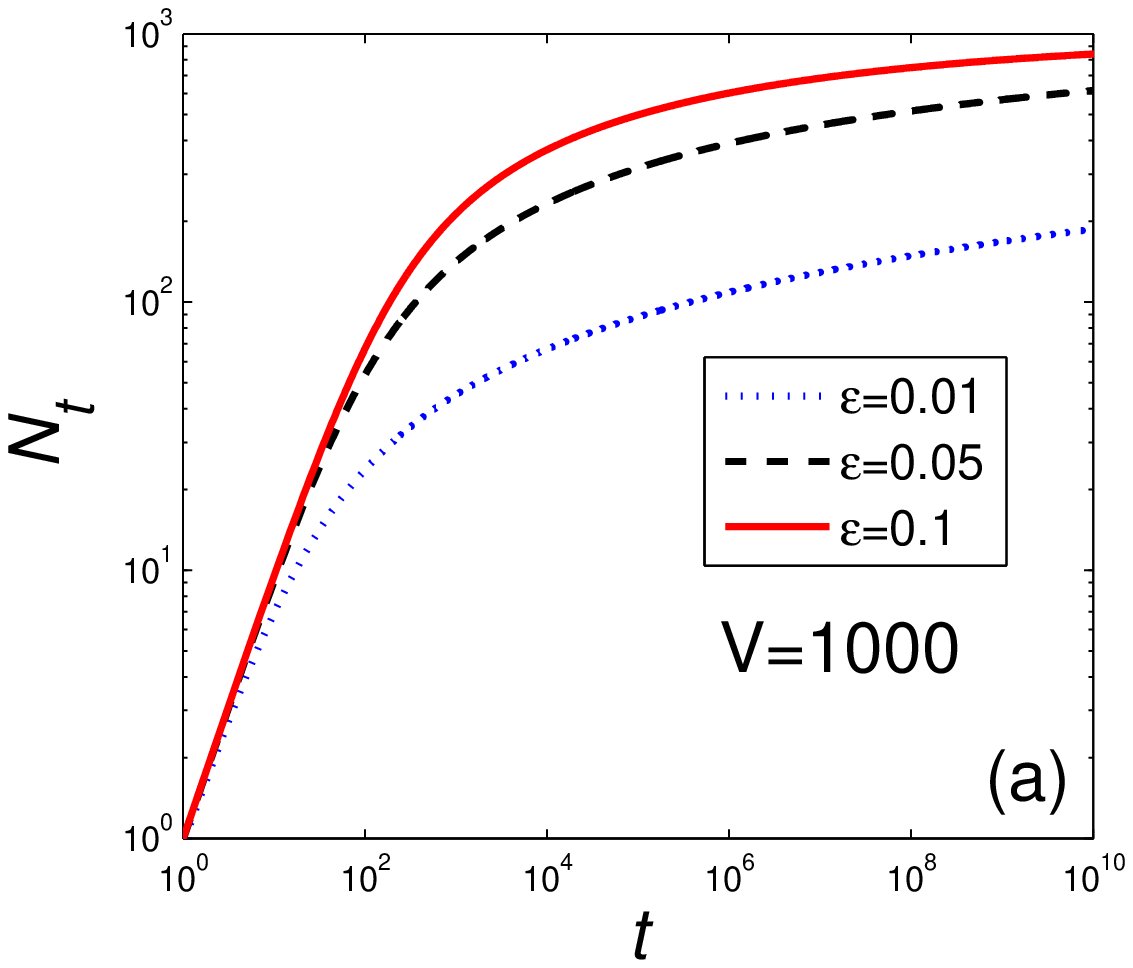}
\includegraphics[width=4.25cm]{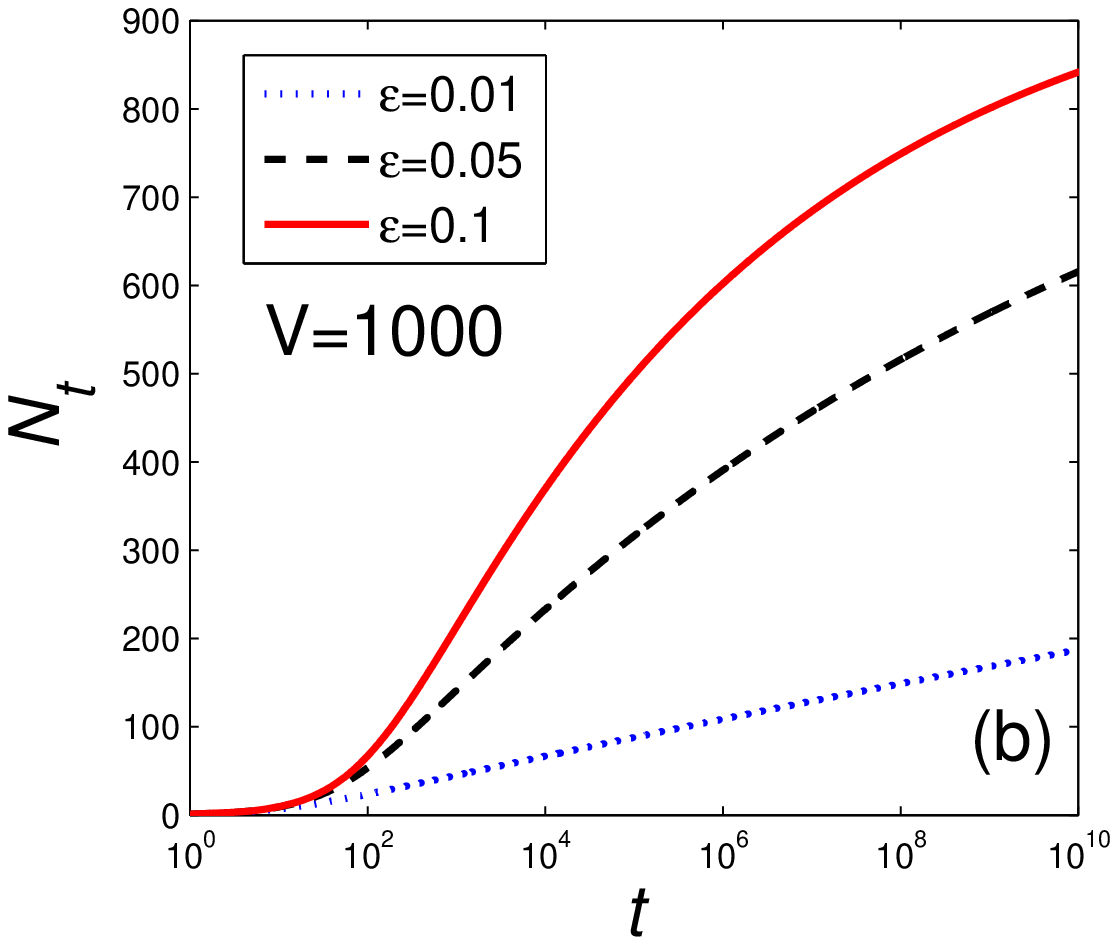}
\includegraphics[width=4.25cm]{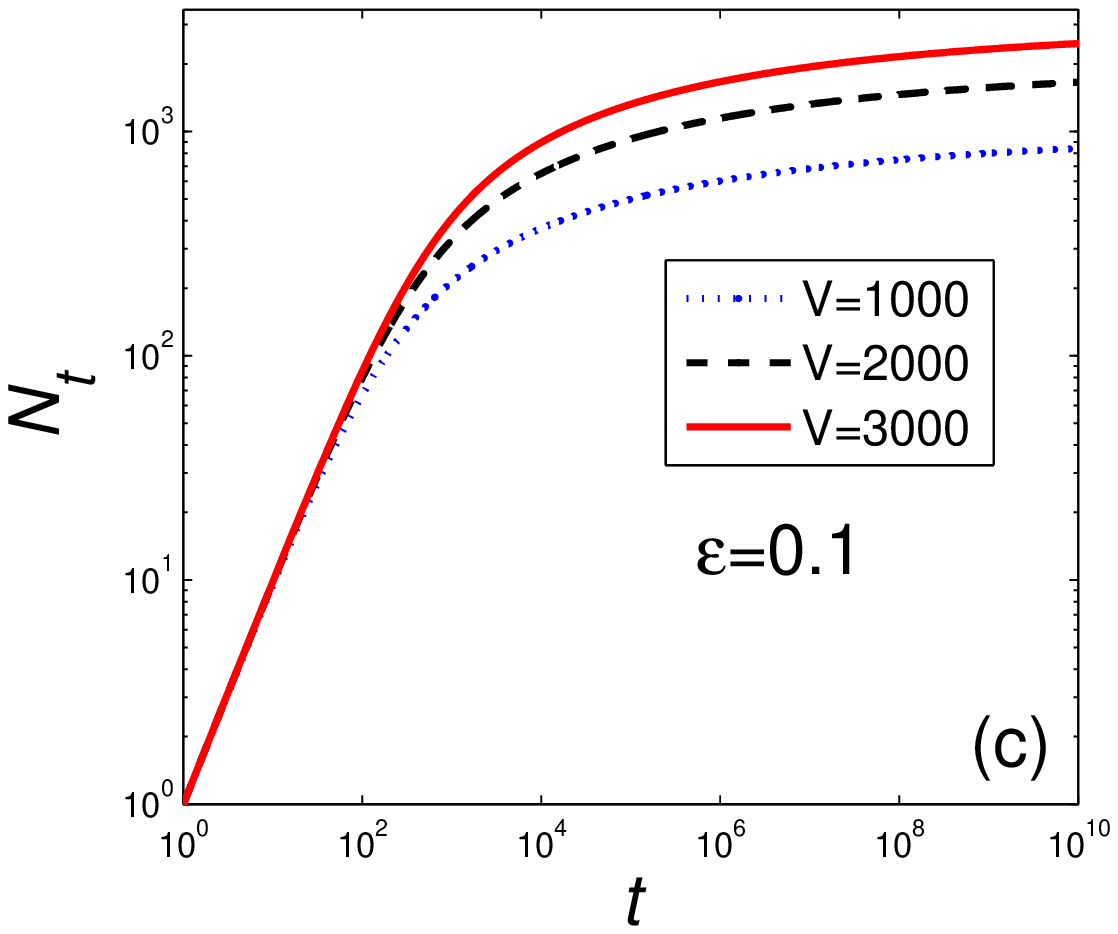}
\includegraphics[width=4.25cm]{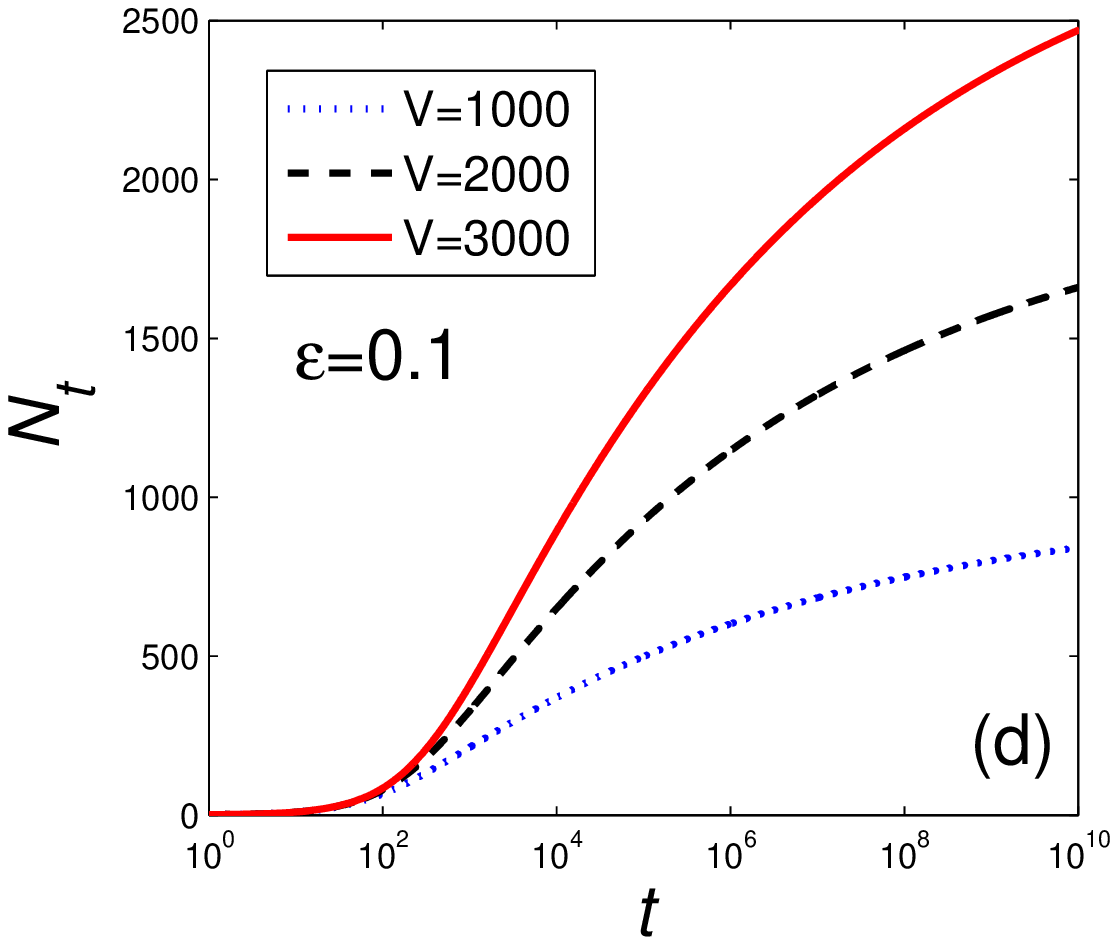}
\caption{(Color online) Growth of the number of distinct characters versus time for different $V$ and $\varepsilon$ according to Eq.~\ref{Nt2}. Plots (a) and (c) are in log-log scale while while (b) and (d) are their corresponding plots in linear-log scale.}\label{numerical}
\end{center}
\end{figure}

This solution embodies three stages of growth of $N_t$. (i) In the very early stage, when $t$ is much smaller than $V\varepsilon$, $(\frac{V\varepsilon}{V\varepsilon+t})^{\varepsilon}\approx{1-\frac{t}{V}}$ and thus $N_t\approx{t}$, corresponding to a short period of linear growth. (ii) When $t$ is of the same order of $V\varepsilon$, if $\varepsilon$ is very small, $N_t$ could be much smaller than $V$. Then Eq. 2 can be approximated as
\begin{equation}
\frac{dN_t}{dt}\approx\frac{V\varepsilon}{V\varepsilon+t},
\end{equation}
leading to a logarithmical solution
\begin{equation}
\label{log}N_t\approx{V\varepsilon{\mathrm{ln}\left(1+\frac{t}{V\varepsilon}\right)}}.
\end{equation}
Indeed, expanding $(\frac{V\varepsilon}{V\varepsilon+t})^{\varepsilon}$ by Taylor series as
\begin{equation}
\left(\frac{V\varepsilon}{V\varepsilon+t}\right)^{\varepsilon}=\sum_{m=0}^{\infty}\frac{1}{m!}\left[\varepsilon\cdot{\mathrm{ln}\left(\frac{V\varepsilon}{V\varepsilon+t}\right)}\right]^m
\end{equation}
and neglecting the high-order terms ($m\geq2$) under the condition $\varepsilon\ll 1$, one can also arrive to the solution Eq.~5. (iii) When $t$ gets larger and larger, $N_t$ will approach to $V$ and thus both $\frac{V\varepsilon}{V\varepsilon+t}$ and $1-\frac{N_t}{V}$ are very small, leading to a very slow growing of $N_t$ according to Eq. 2. These three stages predicted by the analytical solution are in good accordance with the above empirical observations.

Figure 3 reports the numerical results on Eq.~3. In accordance with the analysis, when $t$ is small, $N_t$ grows in a linear form as shown in Fig. 3(a) and 3(c), and from Fig. 3(b) and 3(d), straight lines appear in the middle region, indicating a logarithmical growth predicted by Eq.~5.

\begin{figure}
\begin{center}
\includegraphics[width=4.25cm]{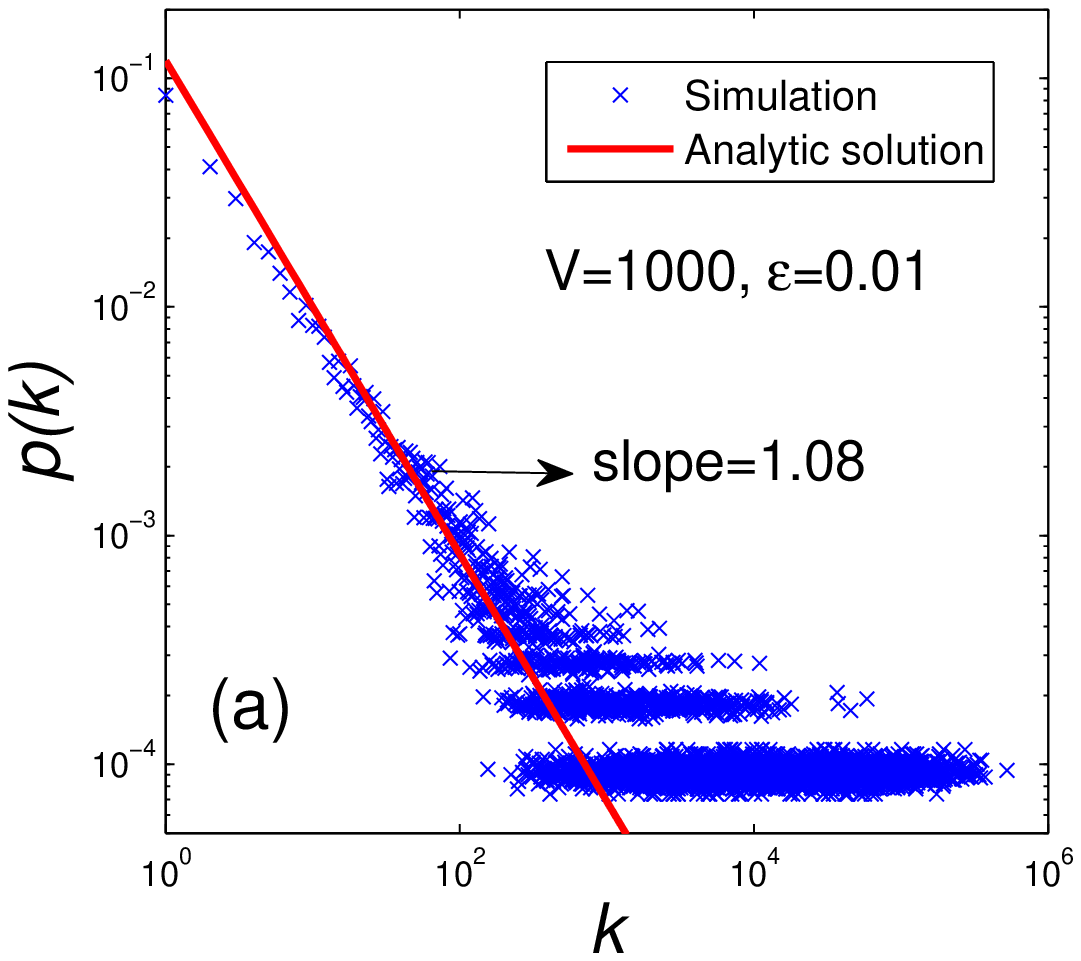}
\includegraphics[width=4.25cm]{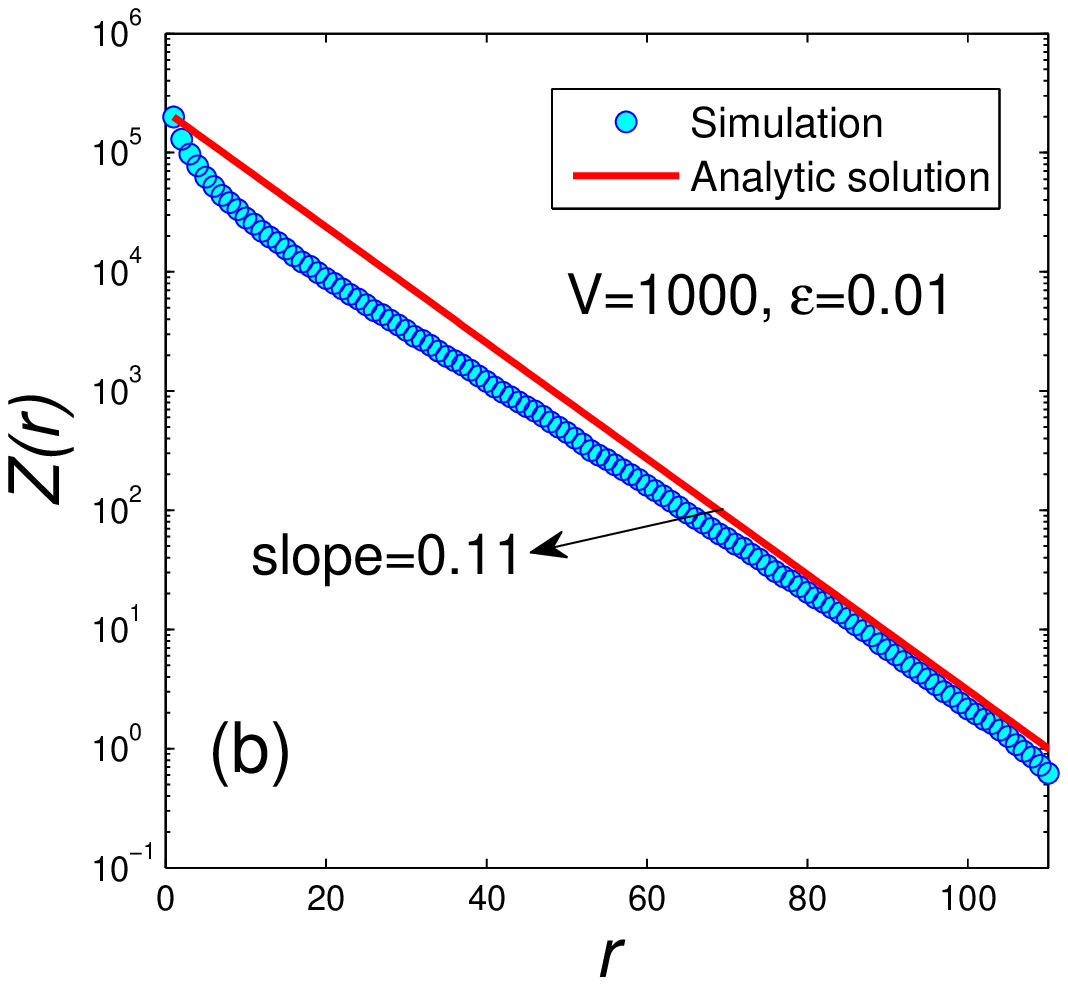}
\includegraphics[width=4.25cm]{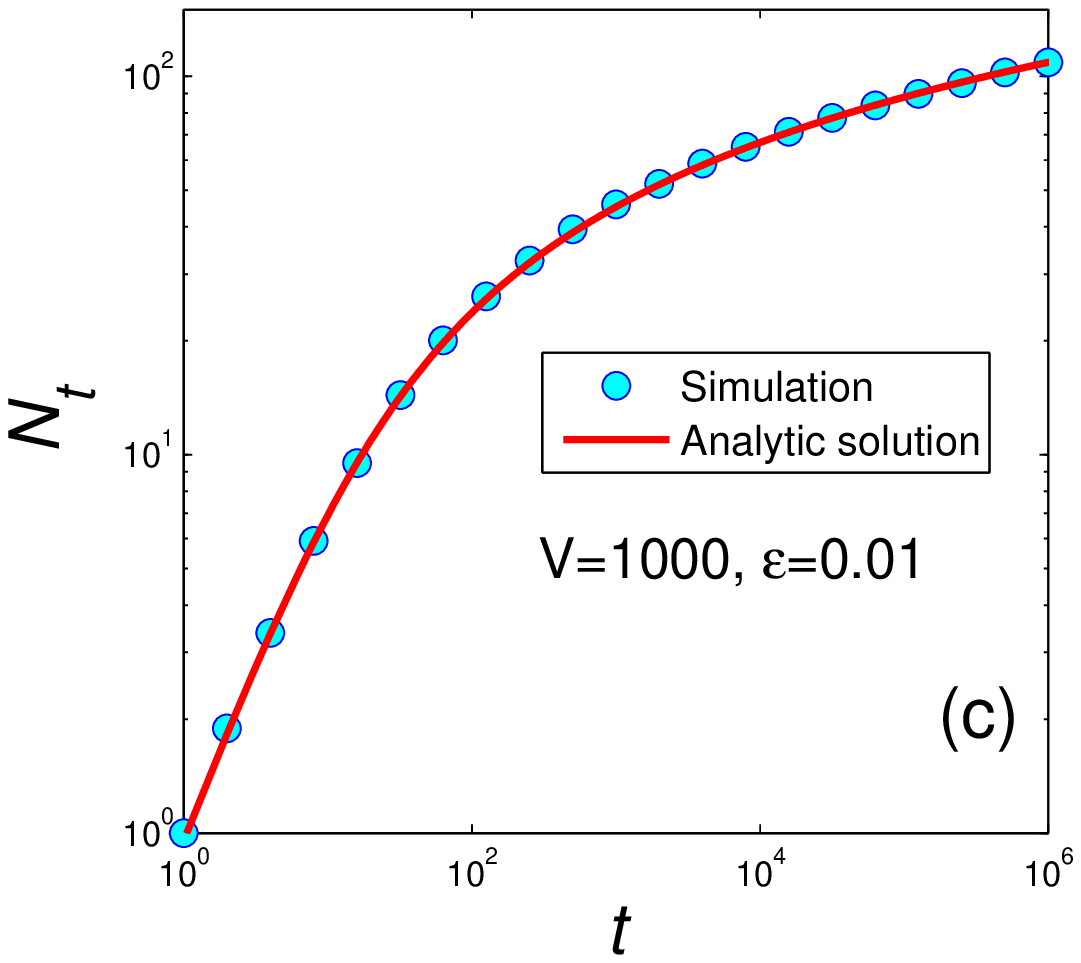}
\includegraphics[width=4.25cm]{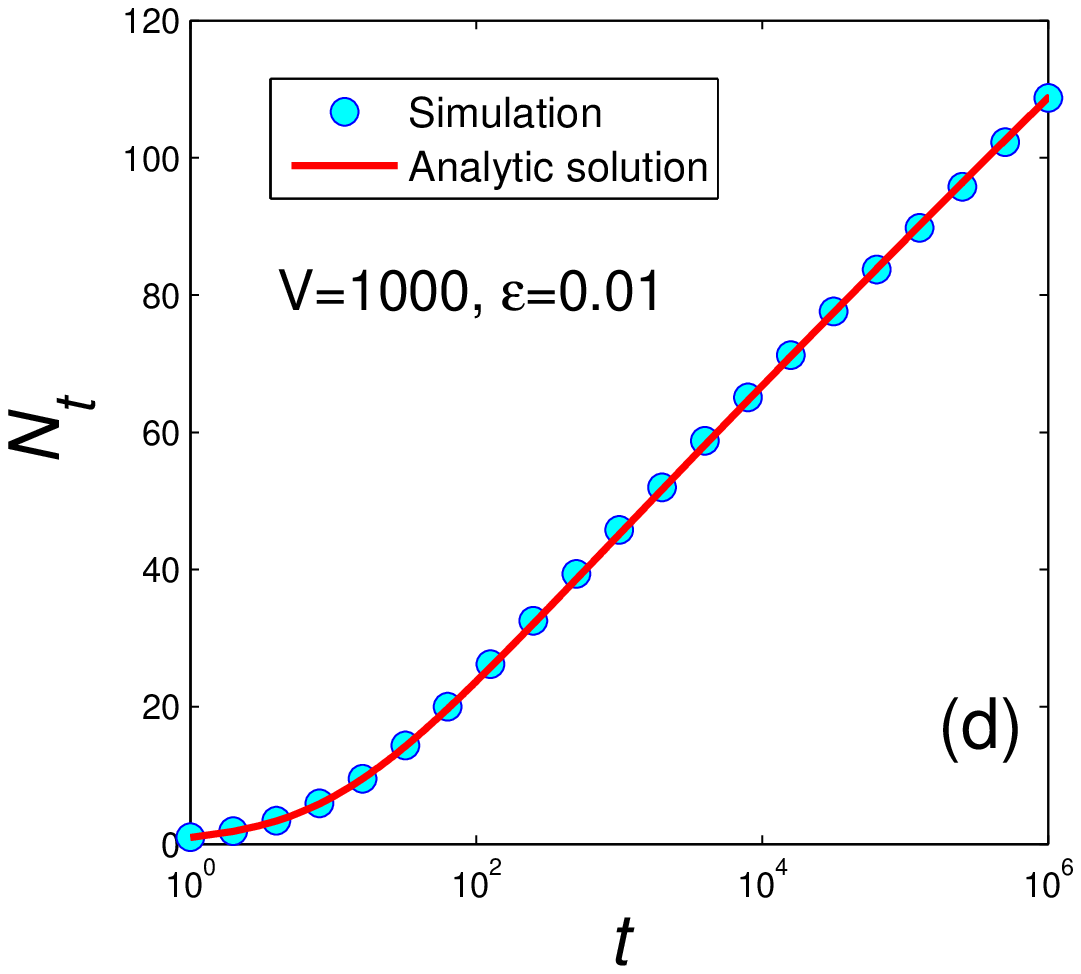}
\caption{(Color online) Comparison between simulations results (blue data points) and analytical solutions (red curves) for typical parameters $V=1000$ and $\varepsilon=0.01$. The subgraphs (a) and (c) are plotted in log-log scale, while (b) and (d) are the same data points to (a) and (b) displayed in log-linear and linear-log scales, respectively. The results are obtained by averaging over 100 independent runs with text length being equal to $10^6$.}\label{numerical}
\end{center}
\end{figure}

Denote by $n(t,k)$ the number of distinct characters that appeared
$k$ times until time $t$, then $n(t,k)=N_tp(k)$. According to the master equations, we have
\begin{equation}
\label{power}
n(t+1,k+1)=n(t,k+1)\left[1-f(k+1)\right]+n(t,k)f(k).
\end{equation}
Substituting Eq.~\ref{fi} into Eq.~\ref{power}, we obtain
\begin{equation}
\label{power2}
N_{t+1}{p(k+1)}=N_t{p(k+1)}\left(1-\frac{k+1+\varepsilon}{V\varepsilon+t}\right)+\frac{N_tp(k)(k+1)}{V\varepsilon+t}.
\end{equation}
Via continuous approximation, it turns to be the following differential equation
\begin{equation}
\frac{dp}{p}=-\left[1+\frac{V\varepsilon+t}{N_t}(N_{t+1}-N_t)\right]\frac{dk}{k+\varepsilon}.
\end{equation}
Substituting $N_{t+1}-N_t=dN_t/dt$ and Eq.~\ref{Nt}, we get the solution
\begin{equation}
\label{power-law}p(k)=B(k+\varepsilon)^{-\left[1+\varepsilon\left(\frac{V}{N_t}-1\right)\right]},
\end{equation}
where $B$ is the normalized factor. The result shows that the character
frequency follows a power-law distribution with exponent changing in time. Considering the finite vocabulary size, in the large limit of $t$, $N_t\rightarrow V$ and thus the power-law exponent, $\beta=1+\varepsilon\left(\frac{V}{N_t}-1\right)$, approaches 1. Under the continuous approximation, the cumulative distribution of character frequency can be written as
\begin{equation}
P(k>k_0)=1-\int_{k_{\min}}^{k_{0}}p(k)dk=1-B\frac{k^{1-\beta}}{1-\beta}|^{k_0}_{k_{\min}},
\end{equation}
where $k_{\min}$ is the smallest frequency. When $\beta\rightarrow1$,
$k^{1-\beta}\approx{1+(1-\beta)\mathrm{ln}k}$, and thus
\begin{equation}
P(k>k_0)=1-B\mathrm{ln}\frac{k_0+\varepsilon}{k_{\min}+\varepsilon},
\end{equation}
where $B\approx\left(\mathrm{ln}{\frac{k_{\max}+\varepsilon}{k_{\min}+\varepsilon}}\right)^{-1}$ according to the normalization condition $\int_{k_{\min}}^{k_{\max}}p(k)dk=1$ and $k_{\max}$ is the highest frequency. According to Eq. 12, there are
$\left(1-B\mathrm{ln}\frac{k+\varepsilon}{k_{\min}+\varepsilon}\right)N_t$ characters having appeared more than $k$ times. That is to say, a character having appeared $k$ times will be ranked at $r=1+\left(1-B\mathrm{ln}\frac{k+\varepsilon}{k_{\min}+\varepsilon}\right)N_t$. Therefore
\begin{equation}
\label{zipf}
Z(r)=k=(k_{\min}+\varepsilon)\mathrm{exp}\left[\frac{1}{B}\left(1-\frac{r-1}{N_t}\right)\right]-\varepsilon,
\end{equation}
and $Z(1)=k_{\max}$, $Z(N_t)=k_{\min}$. In a word, this simple model accounting for the finite vocabulary size results in a power-law character frequency distribution $p(k)\sim k^{-\beta}$ with exponent $\beta$ close to 1 and an exponential decay of $Z(r)$ in the Zipf's plot, which perfectly agree with the empirical observations on Chinese, Japanese and Korean books.

Figure 4 reports the simulation results for typical parameters. The power-law frequency distribution, the exponential decay of frequency in the Zipf's plot and the linear to logarithmic transition in the growth of the distinct number of characters are all clearly observed in the simulation. The simulation results agree very well with the analytical solutions presented in Eq. 3, Eq. 10 and Eq. 13.

Previous statistical analyses about human language overwhelmingly concentrate on Indo-European family, where each language consists of a huge number of words. In contrast, languages consisting of characters, though cover more than a billion people, obtained less attention. These languages include Chinese, Japanese, Korean, Vietnamese, Jurchen language, Khitan language, Makhi language, Tangut language, and many others. Empirical studies here show remarkably different scaling laws of character-formed from word-formed languages. Salient features include an exponential decay of character frequency in the Zipf's plot associated with a power-law frequency distribution with exponent close to 1, and a multi-stage growth of the number of distinct characters. These findings not only complement our understanding of scaling laws in human language, but also refine the knowledge about relationship between the power law and the Zipf's law, as well as the applicability of the Heaps' law. As a result, we should be careful when applying the Zipf's plot for a power-law distribution with exponent around 1, such as the cluster size distribution in two-dimensional self-organized critical systems \cite{Bak1988}, the inter-event time distribution in human activities \cite{Barabasi2005}, the family name distribution in Korea \cite{Kim2005}, species lifetime distribution \cite{Pigolotti2005}, and so on. Meanwhile, we cannot deny a possibly power-law distribution just from a non-power-law decay in the Zipf's plot \cite{Wang2005}.

The currently reported scaling laws can be reproduced by considering finite vocabulary size in a rich-get-richer process. Different from the well-known finite-size effects that vanish in the thermodynamic limit, the effects caused by finite vocabulary size get stronger as the increasing of the system size. Finite choices must be a general feature in selecting dynamics, but not a necessary ingredient in growing systems. For example, also based on the rich-get-richer mechanism, neither the linear growing model \cite{barabasi1999} nor the accelerated growing model \cite{Dorogovtsev2001b} (treating total degree as the text length and nodes as distinct characters, the accelerated networks grow in the Heaps' manner \cite{Lu2010}) has considered such ingredient. The present model could distinguish the selecting dynamics from general dynamics for growing systems.

This work is partially supported by the Swiss National Science
Foundation (Project 200020-132253) and the Fundamental Research Funds for the Central Universities.

\end{document}